# Spin-Orientation Dependent Topological States in Two-Dimensional Antiferromagnetic NiTl$_2$S$_4$ Monolayers

*Jian Liu*[†§], *Sheng Meng*[†§‖⊥], *Jia-Tao Sun*\*[†‡]

[†] Beijing National Laboratory of Condensed Matter Physics, and Institute of Physics, Chinese Academy of Sciences, Beijing 100190, P. R. China

[‡] School of Information and Electronics, Beijing Institute of Technology, Beijing 100081, P. R. China

[§] School of Physical Sciences, University of Chinese Academy of Sciences, Beijing 100049, P. R. China

[‖] Collaborative Innovation Center of Quantum Matter, Beijing, 100190, P. R. China

[⊥] Songshan Lake Materials Laboratory, Dongguan, Guangdong, 523808, P. R. China

\* Address correspondence to: jtsun@iphy.ac.cn (J.T.S.)

\*Tel: +86-10-82649881

**ABSTRACT:** The topological states of matters arising from the nontrivial magnetic configuration provide a better understanding of physical properties and functionalities of solid materials. Such studies benefit from the active control of spin orientation in any solid, which is yet known to rarely take place in the two-dimensional (2D) limit. Here we demonstrate by the first-principles calculations that spin-orientation dependent topological states can appear in the geometrically frustrated monolayer antiferromagnet. Different topological states including quantum anomalous Hall (QAH) effect and time-reversal-symmetry (TRS) broken quantum spin Hall (QSH) effect can be obtained by changing spin orientation in the NiTl$_2$S$_4$ monolayer. Remarkably, the dilated nc-AFM NiTl$_2$S$_4$ monolayer gives birth to the QAH effect with hitherto reported largest number of quantized conducting channels (Chern number $\mathcal{C}$ = –4) in 2D materials. Interestingly, under tunable chemical potential, the nc-AFM NiTl$_2$S$_4$ monolayer hosts a novel state supporting the coexistence of QAH and TRS broken QSH effects with a Chern number $\mathcal{C}$ = 3 and spin Chern number $\mathcal{C}_s$ = 1. This work manifests a promising concept and material realization toward topological spintronics in 2D antiferromagnets by manipulating its spin degree of freedom.

Quantum anomalous Hall (QAH) and quantum spin Hall (QSH) effects are conceptually two milestones of topological states, which have attracted enormous research interests in condensed matter physics and material science.[1-10] Both effects are benefited from the spin degree of freedom coupled with orbital counterparts, endowing solid materials with great potential in spintronics applications.[11-22] The time-reversal-symmetry (TRS) broken QAH and TRS conserved QSH effects manifest the existence of topologically nontrivial states, characterized by the non-zero Chern number $\mathcal{C}$ and odd topological invariant $\mathcal{Z}_2$, respectively.[4-6,9,23] Since the Berry curvature is an odd function, the total QAH conductance of the TRS conserved (Θ) solid materials with a collinear magnetic configuration will vanish. The QSH state consisting of two collinear spin channels of QAH state is described by topological invariant $\mathcal{Z}_2 = (\mathcal{C}_s^\uparrow - \mathcal{C}_s^\downarrow)/2$, where $\mathcal{C}_s$ is the spin Chern number. In this case, $\mathcal{Z}_2$ is equivalent to $|\mathcal{C}_s|$. These novel quantum states of matters originating from the collinear magnetic configuration have lay the foundation of spin dependent topological states in condensed matter physics.

Recently, spin-orientation dependent topological states arising from the noncollinear magnetic configuration in geometrically frustrated antiferromagnetic materials have attracted much attention.[24,25] A number of three-dimensional (3D) antiferromagnetic (AFM) materials with noncollinear *coplanar* magnetic configurations were unexpectedly shown to exhibit the large anomalous Hall effect (AHE) (Figure 1a).[26-30] Furthermore, the noncollinear *noncoplanar* magnetic configuration of solids can acquire an additional phase proportional to the spin chirality $S_1 \cdot (S_2 \times S_3)$, which acts as a fictitious magnetic field yielding the topological Hall effect with nonzero Chern number of $\mathcal{C} = \pm 1$ (Figure 1b).[31,32] These nonvanishing conductivity in the frustrated antiferromagnets not only conflicts with the conventional wisdom that AHE is proportional to the total net magnetization, but also reminds us that novel topological states are related to the complex magnetic configurations therein.[33] The spin-orientation dependent band structures and topological states arise from strong spin-orbit couplings and large magnetocrystalline anisotropy in bulk antiferromagnetic materials. However, two-dimensional (2D) antiferromagnetic monolayer is rarely reported to host spin-orientation dependent topological states to date.

Here we predict by first-principles calculations that the NiTl$_2$S$_4$ monolayer as a geometrically frustrated antiferromagnet has spin-orientation dependent topological states in 2D limit. We found that the noncollinear *noncoplanar* AFM (nc-AFM) state is a QAH insulator with a Chern number $\mathcal{C} = -4$ around the Fermi level significantly different from the state with noncollinear *coplanar* magnetic configurations. This antiferromagnetic monolayer has the largest number of quantized conductance channels hitherto found ever to the best of our knowledge. Interestingly, by tuning the chemical potential, a novel topologically nontrivial phase appears exhibiting both QAH and TRS broken QSH phases simultaneously which arises from the complex spin texture in the momentum space.

As shown in Figure 2a,b, bulk NiTl$_2$S$_4$ crystallizes in a hexagonal lattice with $P\bar{3}m1$ space group (No. 164) and is made of structural units of STl-NiS$_2$-TlS monolayer stacked along the *z* direction with van der Waals (vdW) interlayer separation. The middle NiS$_2$ layer (gray box in Figure 2a) has a CaCl$_2$-type structure and the remaining Tl-S part on both side of NiS$_2$ shares the same structural motif with GaSe.[34-41] As a candidate of spin liquid material, bulk NiTl$_2$S$_4$ is one of the member of large family of triangular lattice antiferromagnets NiX$_2$Y$_4$ (X = Ga, In, Tl, Y = S, Se) compounds.[34-38] As a result, NiTl$_2$S$_4$ monolayer can host a large number of entangled magnetic configurations with coherence beyond the two-spin correlation length at extremely low temperature.[38] Nevertheless, we don't intend to enumerate exhaustedly all the possible magnetic configurations. For the convenience of DFT calculations, we studied the electronic properties of NiTl$_2$S$_4$ monolayer by considering possible magnetic configurations (Supporting Information Figure S2) up to four sublattices (2 × 2 × 1 supercell). [33,42]

We show below that the band structures of NiTl$_2$S$_4$ monolayer show strong dependence on the spin orientation. For example, the t-AFM configuration with 120º *noncollinear coplanar* magnetic configuration (inset of Figure 2c) has the lowest total energy with an indirect band gap of 0.24 eV shown in Figure 2c. In contrast, the nc-AFM NiTl$_2$S$_4$ monolayer with *noncollinear noncoplanar* magnetic configuration (inset of Figure 2d) is a semimetal as shown in Figure 2d and Figure S4c, which can be further fully gapped under 3% tensile strain in the whole Brillouin zone as shown in Figure 2d and Figure S5.

Because of the broken time reversal symmetry, it is expected that the NiTl$_2$S$_4$ monolayer in both t-AFM and nc-AFM states has AHE.[43] To this aim, the anomalous Hall conductivity of NiTl$_2$S$_4$ is calculated by using a tight binding Hamiltonian in maximally localized Wannier functions (MLWF), which is used to fit the first-principles band structures.[44] The Ni $d$, Tl $s$ and S $p$ orbitals were used to constructed the Wannier Hamiltonian amounting to 114 and 152 trial atomic orbitals for t-AFM and nc-AFM configurations respectively. The DFT band structures fitted by Wannier Hamiltonian are discussed in the Figure S10. The transverse Hall conductivity of t-AFM and nc-AFM NiTl$_2$S$_4$ is obtained using the Kubo formula:[45,46]

$$\sigma_{xy}(E_f) = -\frac{1}{2\pi}\frac{e^2}{h}\sum_n \int f(E_f)\Omega^z(k)d^2k, \qquad (1)$$

where $f(E_f)$ is the Fermi occupation distribution, $n$ lables the band number, and $\Omega_n^z$ denotes the $z$ component Berry curvature. The Berry curvature $\Omega^z$ in the above eqs 1 can be determined from the Kubo formula as

$$\Omega^z(k) = -\sum_{m\neq n}\frac{2\mathrm{Im}\langle\psi_{nk}|v_x|\psi_{mk}\rangle\langle\psi_{mk}|v_y|\psi_{nk}\rangle}{(E_{mk}-E_{nk})^2}, \qquad (2)$$

where $\psi_{nk}$ and $E_{nk}$ are the Bloch wave function and eigenvalue of the $n$-th band at $k$ point respectively, and $v_{x(y)}$ is the velocity operator along $x$ ($y$) direction. Specifically, when the Fermi energy $E_f$ lies inside a gap, the quantized Hall conductivity $\sigma_{xy}$ is proportional to Chern number $C$, namely $\sigma_{xy} = C\,e^2/h$.

As shown in Figure 2e, the calculated anomalous Hall conductivity of t-AFM NiTl$_2$S$_4$ without SOC nearly vanishes, which is qualitatively consistent with those of the bulk Mn$_3$Ir and Mn$_3$Pt.[25,26] After including SOC, the calculated anomalous Hall conductivity of t-AFM NiTl$_2$S$_4$ monolayer reaches its maximum value of 4.2 $\Omega^{-1}cm^{-1}$ at −0.10 eV, which can be obtained by hole doping.

We will now show that nc-AFM configuration has the desired topological nontrivial properties, which is yet found in t-AFM state. The Mexican-Hat-Type (MHT) band dispersion (Figure 2d) around the Fermi level is a clear sign of band inversion in the absence of spin-orbit coupling (SOC). We found that the bottom of conduction band (BCB) mainly originates from *s* orbitals of Tl atoms and $p_z$ orbitals of S atoms, while the top of valence band (TVB) is dominated by $(p_x, p_y)$ orbitals of S atoms and $(d_{xz}, d_{yz}, d_{xy}, d_{x^2-y^2})$ orbitals of Ni atoms as shown in Figure S8,S9. Since the contribution of *d* orbitals from Ni atoms is comparable to that of *p* orbitals of S atoms, the MHT band dispersion can be viewed as *p-pd* band inversion. Besides, it is intriguing that the bands in nc-AFM NiTl$_2$S$_4$ without SOC are all doubly degenerate and the spin for each pair of degenerate bands holds the opposite direction but the same magnitude below the Fermi level (the spin eigenvalue in the *x*, *y* and *z* direction as a function of Fermi level is negligible, Supporting Information Figure S7). After the SOC is considered, the two degenerate bands develop a spin splitting without changing the band inversion (Figure 2d and Figure S5b). The momentum dependent band splitting arising from the SOC is very similar with the layered K$_{0.5}$RhO$_2$.[33]

The calculated Hall conductivity $\sigma_{xy}$ of nc-AFM NiTl$_2$S$_4$ monolayer without SOC is displayed as a function of the chemical potential in Figure 2f. We indeed found a clear conductivity plateau with the gap of 37.4 meV around the Fermi level, which is independent on the presence of SOC. The calculated Berry curvature without SOC obtained via eq 2 is plotted in Figure 3a, S11. By integrating the $\Omega(k)$ of the occupied states over the first Brillouin zone (BZ), the Chern number of nc-AFM NiTl$_2$S$_4$ monolayer is −4, in contrast with the Chern number $\mathcal{C} = \pm 1$ of systems with the spin chirality. As mentioned above, the Chern number corresponds to the quantized conductivity $\sigma_{xy} = -4 \cdot e^2/h$. To the best of our knowledge, this is the hitherto reported largest number of quantized conducting channels in antiferromagnetic monolayer.

The bulk-boundary correspondence implies that the topological states with nonzero Chern number $\mathcal{C}$ support $|\mathcal{C}|$ branches of chiral gapless modes localized near the edge.[47,48] To further confirm the existence of the topological states, we calculated the energy bands for the nc-AFM NiTl$_2$S$_4$ nanoribbon of ~80 nanometer (nm) width (Supporting Information Figure S12) using the tight binding model in Wannier basis. The calculated band structures of one-dimensional nanoribbon along the *b* axis is shown in Figure 3b. One can clearly observe that there are eight edge conducting channels of opposite chirality within the band inverted gap,[47,48] confirming the quantized nature of topological edge states in nc-AFM NiTl$_2$S$_4$ monolayer. Figure 3c schematically show the four dissipationless conducting channels propagate parallel along the right (left) edge with positive (negative) group velocity. As mentioned above, bulk bands are all doubly degenerate with their spin exhibiting the opposite direction but the same magnitude. Thus the edge modes are also doubly degenerate and are spin-opposite.

As mentioned above, the nc-AFM NiTl$_2$S$_4$ has Chern number of $\mathcal{C} = -4$, in contrast with the original spin chirality with Chern number of $\mathcal{C} = \pm 1$.[33] To illustrate the physical origin of the large number of quantized channels, the momentum distribution of spin textures are studied shown in Figure 3d. The larger Chern number obtained here stems from the complex spin texture in momentum space (Part VI in Supporting Information). Skyrmion is noncollinear spin textures in which the spin quantization axis changes continuously. For two dimensional systems, the winding number of Skyrmion $Q = \int_{BZ} q(k_x, k_y) dk_x dk_y$ is a topological charge counting the number of times that a unit spin vector $\widehat{S}(k) = S(k)/|S(k)|$ winds around a unit sphere, where

$$q(k_x, k_y) = \frac{1}{4\pi}(\partial_{k_x}\widehat{S} \times \partial_{k_y}\widehat{S}) \cdot \widehat{S}, \tag{3}$$

is the topological charge density in momentum space.[49] We find from the real-spin textures and the topological charge density, as shown in Figure 3d that each $K$ valley for one VB consists of five half-skyrmions of $Q = -\frac{1}{2}$ (representing one half bloch-skyrmion and four half anti-skyrmions) and three half bloch-skyrmions of $Q = \frac{1}{2}$. Thus the net half-skyrmions give rise to a winding number $Q = -1$ at each $K$ valley and a total winding number $Q = -2$ for one VB. Considering the fact that the two degenerate VBs are spin-opposite shown above, their Chern numbers have the same sign, then generating a total Chern number of $C = -4$.

Besides the nontrivial gap around the Fermi level, another small gap at –0.52 eV below the Fermi level is shown in Figure 4a. The main contribution of the Berry curvature $\Omega(k)$ obtained by eq 2 comes from two degenerate valleys $K$ and $K'$ (Figure 4a). By integrating the $\Omega(k)$ over the whole BZ, a relatively larger Chern number $C = 3$ is obtained. Since the spatial inversion symmetry of NiTl$_2$S$_4$ monolayer is preserved, one may question the existence of QSH state considering the fact that the bands are all doubly degenerate below the Fermi level. To demonstrate this point, we calculated the spin Berry curvature $\Omega_s^z(k)$ by replacing the velocity operator $\hat{v}_x$ with the spin current operator $\hat{j}_x = \{\hat{v}_x, \hat{s}_z\}$ in eq 2, where $\hat{s}_z$ is the spin operator.[50-52]

The spin Chern number $C_s$ is obtained in the same way as the Chern number $C$ as shown before. Figure 4b shows the calculated spin Berry curvature $\Omega_s^z(k)$ in 2D momentum space. The calculated spin Chern number $C_s = 1$ indicates that nc-AFM NiTl$_2$S$_4$ has the time-reversal-symmetry-broken QSH effect at −0.52 eV. This state is further verified by the calculated Wannier charge center (WCC) (Figure S13). This is different from the QAH state of nc-AFM NiTl$_2$S$_4$ with $C = -4$ and $C_s$=0 around the Fermi level . Figure 4c,d show the corresponding surface states on the left and right surface for the (010) surface BZ, respectively. The emergence of three topologically protected gapless edge states near −0.52 eV is qualitatively consistent with the quantized Hall conductivity of $3\,e^2/h$.

In conclusion we have demonstrated that the spin-orientation dependent topological states are feasible in the two-dimensional antiferromagnetic NiTl$_2$S$_4$ monolayer. The nc-AFM NiTl$_2$S$_4$ monolayer exhibits intriguing spin chirality hosting the QAH phase. This state has a very large number of quantized conductance channels with a gap size of 37.4 meV at the Fermi level. Moreover, we reports the first 2D antiferromagnetic monolayer exhibiting both QAH and TRS broken QSH effects at −0.52 eV below the Fermi level. As the same family of NiTl$_2$S$_4$, the band structures of nc-AFM NiX$_2$Y$_4$ (X=Ga, In, Y=S, Se) shown in Figure S4 are expectedly very similar with that of NiTl$_2$S$_4$ around the Fermi level. The above discussion on the spin-orientation dependent topological states is thus applicable to the same material family. Our work demonstrates spin orientation could serve as a new degree of freedom of electrons enabling to manipulate the topological states of two-dimensional magnetic monolayers.

Finally, we comment on the magnetic configuration and the corresponding topological states in monolayer $NiTl_2S_4$ and provide some clues to the experiment feasibility of our prediction. Bulk $NiGa_2S_4$ is a spin liquid material with numbers of magnetic configurations depending on the temperature. Previous experiments have confirmed the existence of noncollinear coplanar AFM configuration in bulk $NiGa_2S_4$.[38] In principle, the phase transition with different magnetic configurations can be realized by controlling the temperature.[28] As the same family of $NiGa_2S_4$, however, the electronic structures and magnetic configuration of bulk $NiTl_2S_4$ has never been reported before, which is a new system to certain extent. The topological states with different magnetic configurations in 2D $NiTl_2S_4$ can be, in a same way, measured using the transport experiment as mentioned in bulk $Mn_3Pt$, $Mn_3Sn$ and $NiGa_2S_4$.[27,38,53,54] The phase transition between different magnetic configuration in 2D $NiTl_2S_4$ can be studied by transport experiment with exactly controlled temperature.

**ASSOCIATED CONTENT**

The supporting information is available free of charge on the ACS Publications website at DOI: XXX.XXX/XXXXXX

A brief description of the methods; the magnetic configurations and the corresponding electronic structures of $NiTl_2S_4$ monolayer; the fitted band structure by maximally localized Wannier functions; the momentum distributions of the spin Berry curvature at different chemical potentials; the real spin texture in momentum space; the on-site Coulomb interaction U effect on the topological states. (PDF)

AUTHOR INFORMATION

Corresponding Authors

*E-mail: jtsun@iphy.ac.cn (J.T.S.)

ORCID

Jia-Tao Sun: 0000-0003-3519-352X

Sheng Meng: 0000-0002-1553-1432

Notes

The authors declare no competing financial interest.

ACKNOWLEDGMENTS

We thank Hongming Weng for fruitful discussion. This work was partially supported by the National Key Research and Development Program of China (2016YFA0202300), the National Natural Science Foundation of China (Grant Nos. 61725107, 51572290 and 11334006), and the Chinese Academy of Sciences (Grant Nos. XDB06, XDB30000000 and XDB07030100). Most of the calculations were carried out on the Tianjin Supercomputing Center.

**Figures**

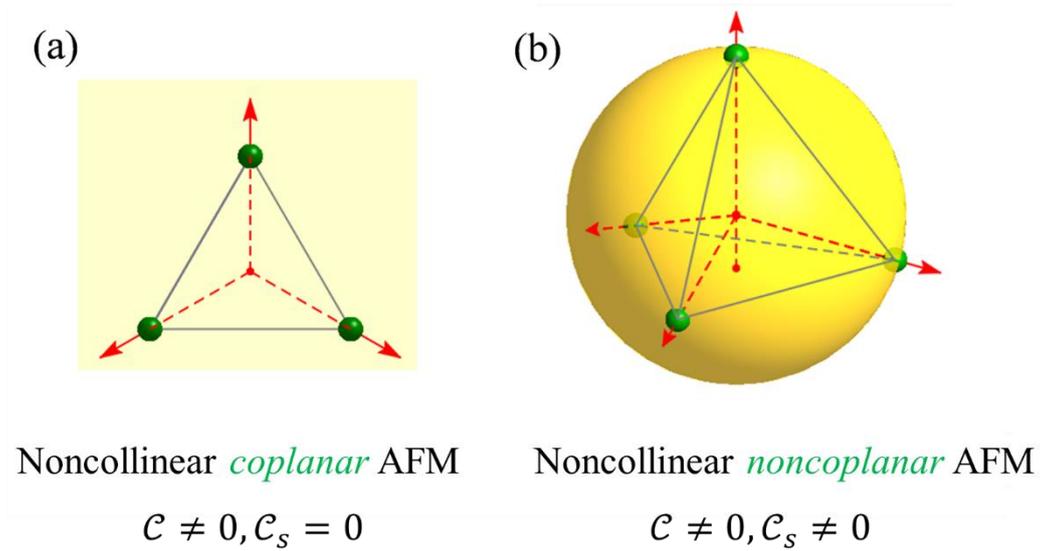

**Figure 1.** Schematics of the spin orientation dependent QAH and QSH effect associated with noncollinear *coplanar* (a), and noncollinear *noncoplanar* (b) antiferromagnetic state. The Chern number and spin Chern number is denoted by $\mathcal{C}$, $\mathcal{C}_s$ respectively.

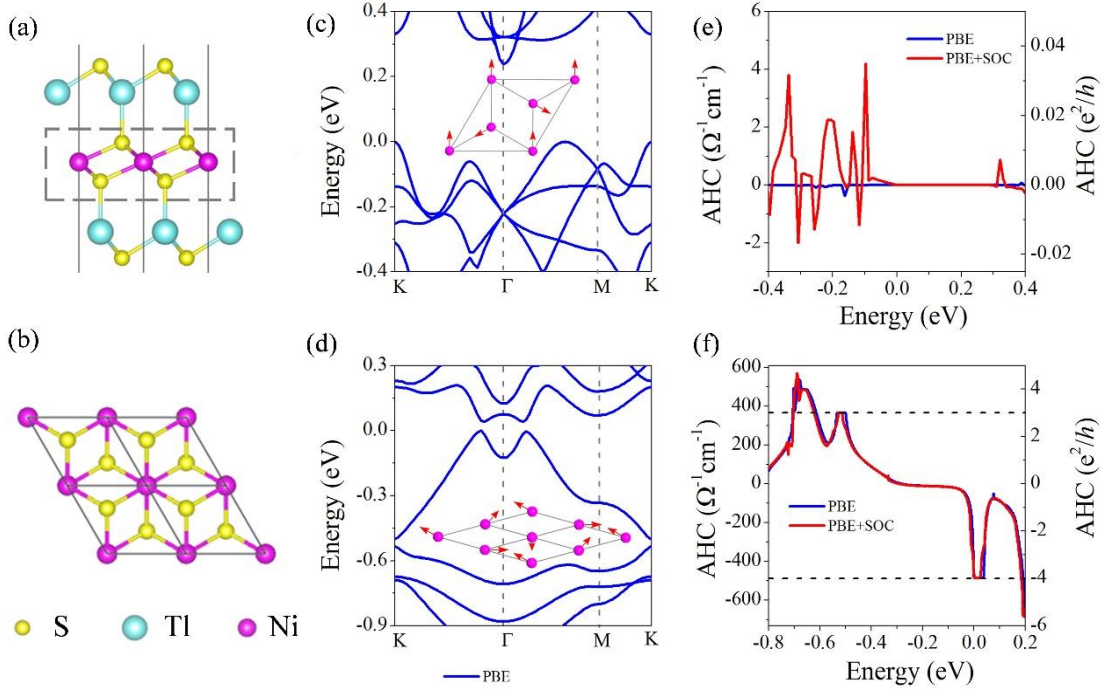

**Figure 2.** The spin-orientation dependent band structures and anomalous Hall effect of antiferromagnetic NiTl$_2$S$_4$ monolayer. (a) The side view of NiTl$_2$S$_4$ monolayer. (b) The top view of the middle subunit NiS$_2$ shown by dashed grey line in (a). (c) Band structure of the noncollinear *coplanar* antiferromagnetic (t-AFM) state without SOC. The inset is the schematic diagram of t-AFM. (d) Band structure of the 3% dilated nc-AFM NiTl$_2$S$_4$ monolayer associated with noncollinear *noncoplanar* antiferromagnetic state (inset) without considering the SOC. (e) Anomalous Hall conductivity of t-AFM NiTl$_2$S$_4$ monolayer without (blue lines) and with (red lines) SOC. (f) Anomalous Hall conductance of nc-AFM NiTl$_2$S$_4$ monolayer without (blue lines) and with (red lines) SOC. The red arrows in (c) and (d) represent the spin orientation of each Ni atom (magneta spheres). The Fermi level is shifted at the valence band maximum for the sake of comparison.

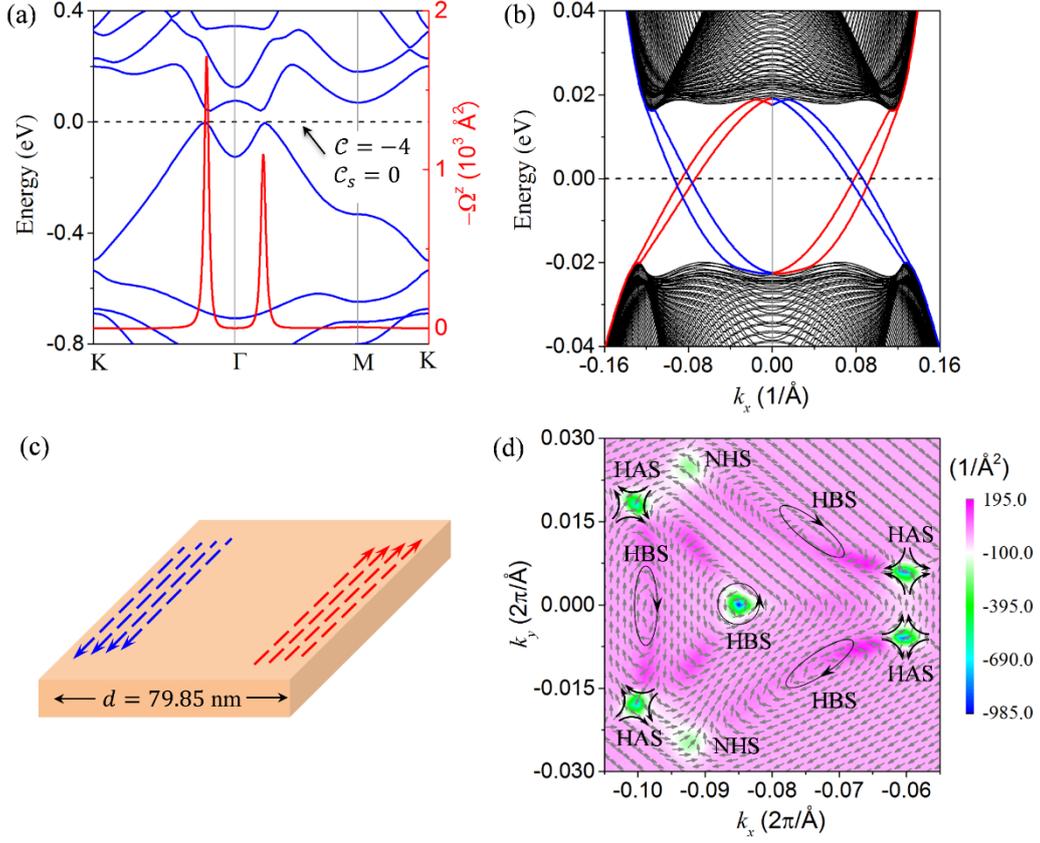

**Figure 3.** The calculated Berry curvature and topologically nontrivial edge states of nc-AFM NiTl$_2$S$_4$ monolayer. (a) Band structure (blue curves) and the distribution of Berry curvature $-\Omega^z(k)$ (red curves) along high symmetric momentum path. (b) Electronic band structure of the nc-AFM NiTl$_2$S$_4$ nanoribbon with the width of 79.85 nm. (c) Schematics of the propagation direction of spin-opposite edge modes in real space. (d) Spin texture $(S_x, S_y)$ for one of two degenerate valence bands around one K valley. The topological charge density is in unit of $(1/\text{Å})^2$. The length and direction of the gray arrows represent the in-plane component of the normalized spin texture. The black arrows schematically depict the vortex direction. Each blue and magenta region correspond to a half-skyrmion, half bloch-skyrmion (HBS) and half anti-skyrmion (HAS). NHS denotes the spin texture without forming half-skyrmion.

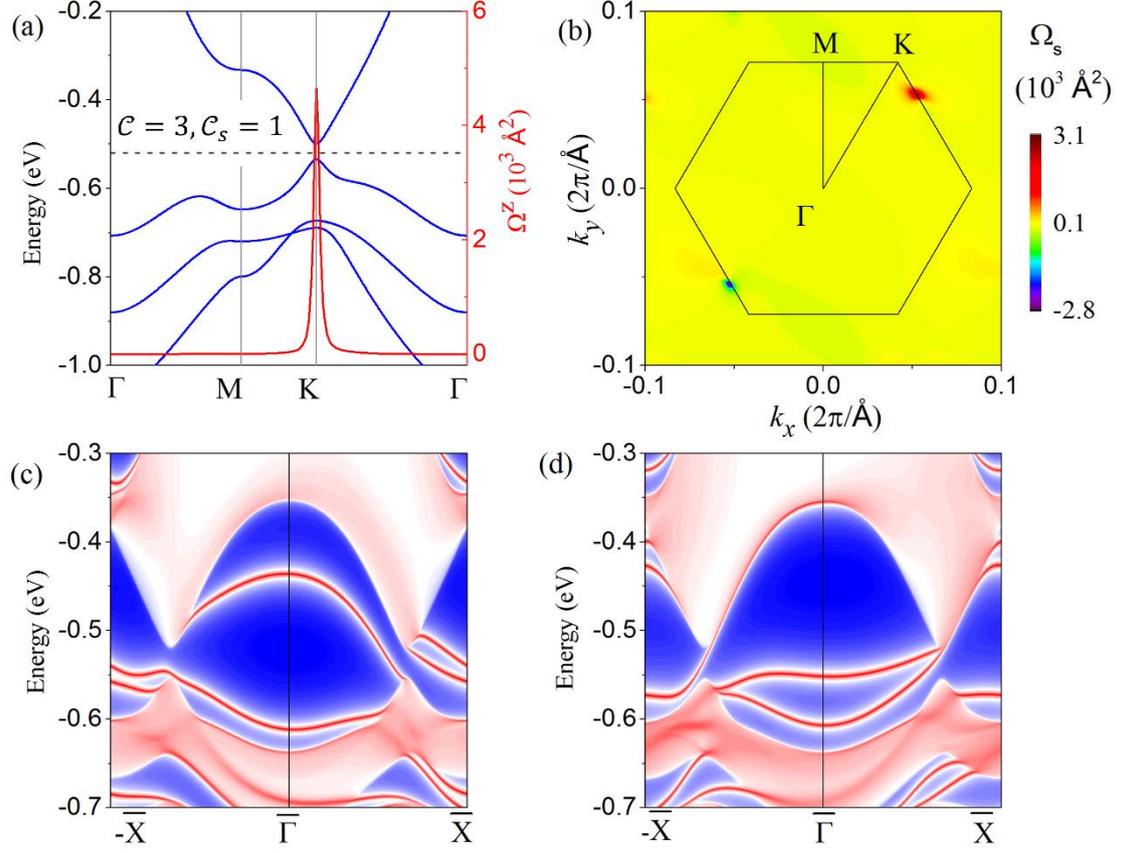

**Figure 4.** Topological states of 3% dilated nc-AFM NiTl$_2$S$_4$ monolayer when the chemical potential lies near 0.52 eV below the Fermi level. (a) Band structure (blue curves) and the distribution of Berry curvature (red curves) along high symmetry points. (b) 2D momentum distribution of the spin Berry curvature $\Omega_s(k)$. The surface states on the left (c) and right (d) sides of the nc-AFM NiTl$_2$S$_4$ nanoribbon respectively.

# TOC

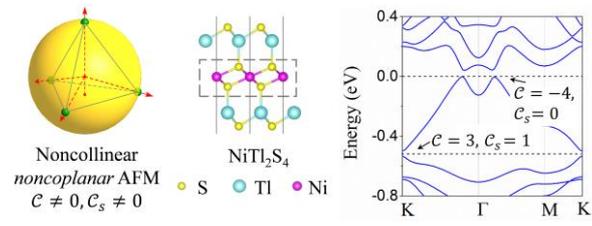